# Exploring the Structural Stability, Electronic and Thermal Attributes of synthetic 2D Materials and their Heterostructures

Ghulam Hussain[a,*], Mazia Asghar[b], Muhammad Waqas Iqbal[b], Hamid Ullah[b,*], Carmine Autieri[a]

[a]International Research Centre MagTop, Institute of Physics, Polish Academy of Sciences, Aleja Lotników 32/46, PL-02668 Warsaw, Poland

[b]Department of Physics, Riphah International University, Campus Lahore, Pakistan

E-mail: ghussain@ifpan.edu.pl, hamid.ullah@riphah.edu.pk

## ABSTRACT

Based on first-principles calculations, we have investigated the structural stability, electronic structures, and thermal properties of the monolayer $XSi_2N_4$ (X= Ti, Mo, W) and their lateral (LH) and vertical heterostructures (VH). We find that these heterostructures are energetically and dynamically stable due to high cohesive and binding energies, and no negative frequencies in the phonon spectra. The $XSi_2N_4$ (X= Ti, Mo, W) monolayers, the $TiSi_2N_4/MoSi_2N_4$-LH, $MoSi_2N_4/WSi_2N_4$-LH, and $MoSi_2N_4/WSi_2N_4$-VH possess a semiconducting nature with an indirect band gap ranging from 0.30 to 2.60 eV. At room temperature, the $C_v$ values are found to be between 100 and 416 JK$^{-1}$mol$^{-1}$ for the monolayers and their heterostructures, suggesting the better ability to retain heat with respect to transition metal dichalcogenides. Our study unveils the excellent attributes of $XSi_2N_4$ 2D monolayers and their heterostructures, proposing them as potential candidates in nanoelectronics and thermoelectric applications.

1. **Introduction**

Due to the monolayer limit, the two-dimensional (2D) materials have distinct physical properties and thus are employable in a wide range of device applications [1-9]. Graphene, the first 2D material, has been widely studied since its discovery [3, 10]. It was expected to be a suitable layered structure for the new generation of nanoelectronic devices thanks to its high carrier mobility. However, graphene's zero band gap reminds us that it cannot be effective in many device applications [11-13]. On the other hand, transition metal dichalcogenides (TMDCs) are explored to exhibit tunable band gaps, but on account of practical applications, their relatively low carrier mobilities cannot be neglected [1, 14, 15]. For instance, $MoS_2$ possesses carrier mobility of approximately 200 $cm^2V^{-1}s^{-1}$ for holes and 72 $cm^2V^{-1}s^{-1}$ for electrons [14], which are much lower in magnitude than that of graphene and even much smaller than Si (480 $cm^2V^{-1}s^{-1}$ for hole and 1350 $cm^2V^{-1}s^{-1}$ for electron) [14, 16]. In recent years, from the fundamental and applicative point of view, we have assisted to tremendous efforts in the investigation of the topological and magnetic properties of the 2D materials and quasi-2D materials [17-25].

Recently, the discovery of novel 2D materials family, $XA_2Z_4$ (X=Transition metal, A=IV element, Z=V element) [26] has attracted broad consideration due to the outstanding properties they demonstrate. The first member of this assembly is the $MoSi_2N_4$ that was successfully produced via chemical vapor deposition (CVD) process with a large area of size 15 mm × 15 mm. This semiconducting material with a septuple-atomic-layer configuration (N-Si-N-Mo-N-Si-N) can be regarded as a $MoN_2$ layer sandwiched in two SiN layers, exhibiting an indirect band gap of almost 1.94 eV. The monolayer of $MoSi_2N_4$ depicts an elastic modulus that is roughly four times larger than $MoS_2$, and the carrier mobilities of holes and electrons are also four to six times that of a single layer of $MoS_2$. Moreover, a new family, namely $XA_2Z_4$ (X=Transition metal, A=IV element,

Z=V element) is anticipated by density functional theory (DFT) calculation. So far, several studies have been carried out describing the exceptional features of these materials [27-34].

Designing vertically stacked or laterally stitched two-dimensional heterostructures between 2D materials is an effective technique to unravel new features and extend their applications to electronics, detectors, electroluminescence, and photovoltaics [35-41]. Several experimental techniques such as physical vapor transport, CVD, plasma-assisted deposition, thermal decomposition, and vapor phase growth method have been utilized to synthesize the lateral heterostructures (LH) and vertical heterostructures (VH) [41-47]. Recently, the vertically stacked bilayer of $MoSi_2N_4$ have been studied using first-principles calculations, which reports the effect of strain on the electronic band gap, the bilayer $MoSi_2N_4$ showed a decrease in the band gap as the vertical strain was increased; and at around 22% strain, the bilayer revealed a semiconducting to metal transition [31]. Also, the band gap decreases as a function of the electric field for both the $MoSi_2N_4$, and $WSi_2N_4$ bilayers, respectively [28]. Likewise, the vertical heterostructures of $MoSi_2N_4$ and $MoGe_2N_4$ have been reported, where the electronic properties and ferroelectricity are investigated [29, 30]. Nonetheless, the van der Waals heterostructures of this class are not explored thoroughly and demand a detailed study. In addition, the laterally stitched heterostructures of this novel family are not reported so far.

In this work using first-principle calculations, we carry out a detailed study of $XSi_2N_4$ (X= Ti, Mo, W) 2D monolayers and their lateral and vertical heterostructures such as $TiSi_2N_4/MoSi_2N_4$ and $MoSi_2N_4/WSi_2N_4$. All the compounds are investigated in their 2H crystal structure. The cohesive energies and the absence of imaginary phonon modes in the phonon dispersions exposed the structural stability of $XSi_2N_4$ systems and their lateral and vertical heterostructures. Since the electronic properties calculated by DFT are usually in poor agreement with the experiments, principally the band gap, the Hybrid functional (HSE06) is applied to accurately approximate the electronic band structures. Both, the density of states and electronic band structures of $XSi_2N_4$ 2D materials and their LHs and VHs indicated trivial indirect band gaps. In contrast, a semiconducting to metallic transition is observed in the vertical heterostructure of $TiSi_2N_4/MoSi_2N_4$. Compared to parent $XSi_2N_4$ materials, the lateral and vertical heterostructures show the highest free energy and heat capacity values. Our theoretical investigations could pave the path for designing novel nanoelectronic devices based on $XSi_2N_4$ and their hetero-systems.

## 2. Methodology

We performed first-principles calculation via the Vienna ab initio simulation package (VASP) [48, 49] using density functional theory (DFT). Both the Heyd-Scuseria-Ernzerhof (HSE06) and the generalized gradient approximation with PBE form [27] and [28] are adopted to calculate the electronic properties [50, 51]. For relaxation, an energy cutoff of 500 eV and a mesh of $15 \times 15 \times 1$ k-points were chosen. The periodic structures in the perpendicular direction of 2D material and heterostructure were separated with a vacuum layer of approximately 20 Å, which would avoid spurious interactions. Our calculations were converged by setting the convergence criteria to 0.01 eV/Å and energy $10^{-7}$ eV. To study the dynamical stability of $XSi_2N_4$ monolayers and that of LHs and VHs, PHONOPY code using a 4×4×1 supercell was employed for calculating the phonon band structures [52]. Also, the thermal properties such the heat capacity, free energy and entropy were explored using phonon calculations.

## 3. Results and discussions

*Structural stability*

The optimized structures for $XSi_2N_4$ (X=Ti, Mo, W) monolayers and their heterostructures are visualized in Fig. 1, where the (a), (b) and (c) panels depict the monolayer, lateral heterostructure (LH) and vertical heterostructures (VH), respectively. It can be seen that $XSi_2N_4$ (X=Ti, Mo, W) possesses a honeycomb structure consisting of Ti/Mo/W, Si and N atoms. The side view illustrates that this structure is stacked by seven atomic layers of N-Si-N-X-N-Si-N that exhibits a sandwich structure of $MoN_2$ layer by two SiN bilayers. We noticed the optimized lattice constants for $XSi_2N_4$ (X=Ti, Mo, W) monolayers to be $a = b = 2.932$, $2.909$ and $2.915$ Å, respectively. These lattice constants are in perfect agreement with the calculated lattice parameters in the literature [26, 27, 53, 54]. Here, we consider the lateral AA- stitching for the LHs (Fig. 1b) and vertical AB-stacking for the VHs (Fig. 1c), the heterostructures obtained are $TiSi_2N_4/MoSi_2N_4$-LH, $TiSi_2N_4/MoSi_2N_4$-VH, $MoSi_2N_4/WSi_2N_4$-LH, and $MoSi_2N_4/WSi_2N_4$-VH. In the AA-stacking of VHs, all atoms i.e. the X, Si and N atoms of one monolayer of $XSi_2N_4$ are located exactly above the X, Si and N atoms of the other $XSi_2N_4$ layer. Whereas, in AB-stacking, all the atoms in the one $XSi_2N_4$ layer are staggered with respect to another $XSi_2N_4$ monolayer, as depicted in Fig. 1c. Based on total ground state energy calculations, the ground state is AB stacking [26, 28, 29].

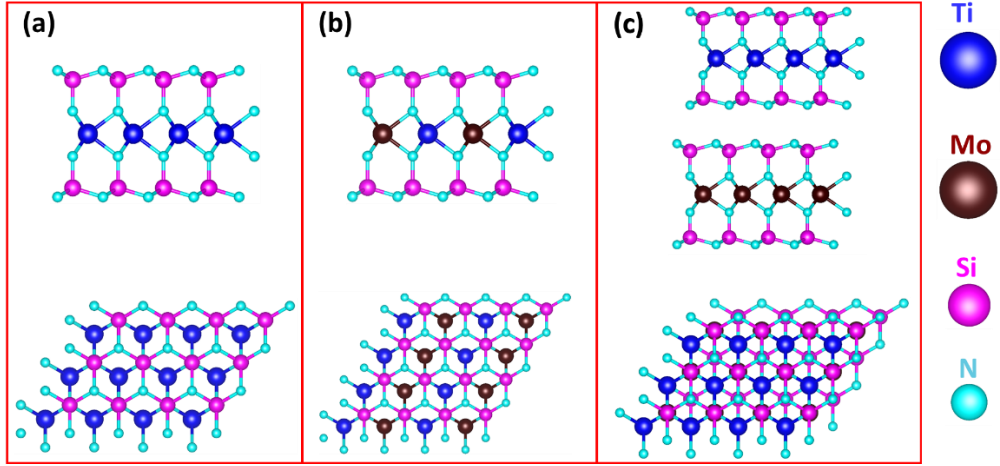

**Figure 1** Side and top views of the atomic structures of (a) $XSi_2N_4$ monolayer, (b) lateral heterostructure (LH), and (c) vertical heterostructure (VH).

In order to check the stability of $XSi_2N_4$ monolayers and that of LHs and VHs, we used the total ground state energies to compute the cohesive energies. In addition, the phonon dispersions are calculated to determine their dynamical stability, which can enable the use of the studied systems in the device application. In the monolayer case of $XSi_2N_4$, the cohesive energy is calculated by [55, 56];

$$E_c = \frac{1E_X + 2E_{Si} + 4E_N - E_{XSi_2N_4}}{1 + 2 + 4}$$

Here, $E_X$, $E_{Si}$, $E_N$ and $E_{XSi_2N_4}$ are total ground state energies of isolated X, Si, N atoms and monolayer $XSi_2N_4$, respectively. The cohesive energies for $TiSi_2N_4$, $MoSi_2N_4$ and that of $WSi_2N_4$ are evaluated to be 5.562, 5.589 and 5.653 eV/atom, respectively. By comparing these values with that of graphene (7.46 eV/atom), $MoS_2$ (4.98 eV/atom) and phosphorene (3.30 eV/atom) [57-59], we noted that they are smaller than graphene but larger than $MoS_2$, which indicate the better stability. For the heterostructures, the binding energies are estimated to determine the stability, it is given by [60];

$$E_b = E_{heterostructure} - E_{X^1Si_2N_4} - E_{X^2Si_2N_4}$$

Where $E_{heterostructure}$, denotes the total energy of the heterostructure while $E_{X^1Si_2N_4}$, $E_{X^2Si_2N_4}$ represent the ground state energies of the two monolayers involved in the formation of LH or VH. The binding energies for $TiSi_2N_4/MoSi_2N_4$-LH, $TiSi_2N_4/MoSi_2N_4$-VH, $MoSi_2N_4/WSi_2N_4$-LH, and $MoSi_2N_4/WSi_2N_4$-VH are -120.175 eV, -45.06 meV, -125.465 eV and -50.082 meV, respectively.

To examine the dynamical stability of $XSi_2N_4$ (X=Ti, Mo, W) monolayers and that of

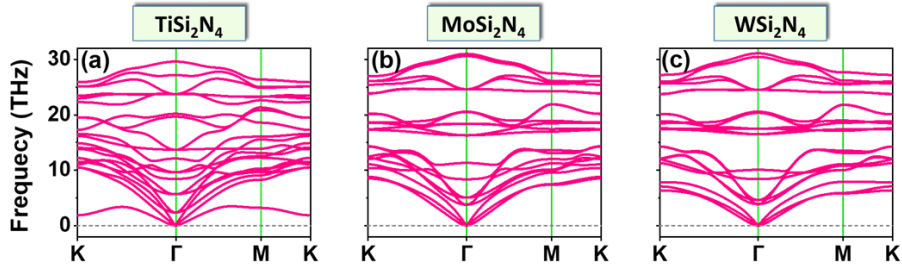

**Figure 3** Phonon bandstructures for $XSi_2N_4$ (X=Ti, Mo, W) monolayers revealing no imaginary modes.

heterostructures, we carried out the phonon dispersion calculations as shown in Fig. 2 and 3. The

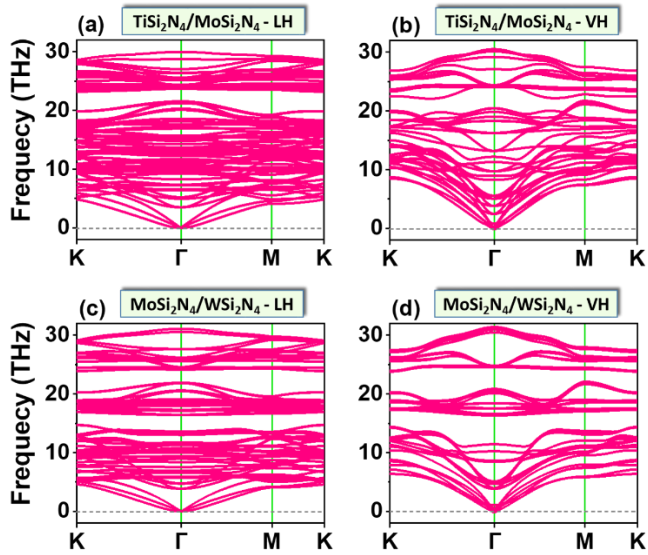

**Figure 2** Calculated phonon dispersion curves of the lateral and vertical heterostructures for (a) $TiSi_2N_4/MoSi_2N_4$-LH, (b) $TiSi_2N_4/MoSi_2N_4$-VH, (c) $MoSi_2N_4/WSi_2N_4$-LH and (d) $MoSi_2N_4/WSi_2N_4$-VH.

finite difference method is employed to calculate the phonon spectra using the Phonopy code [52]

with VASP [61] as a calculator. From the phonon dispersion curves shown in Fig. 2, it is confirmed that all the three $XSi_2N_4$ (X=Ti, Mo, W) monolayers are dynamically stable due to the absence of imaginary modes. Similarly in Fig. 3, the calculated phonon band spectra along the Brillouin zone's high symmetry directions (K-Γ-M-K) for both the lateral and vertical heterostructures reveal no imaginary frequency at the Γ-point, indicating the dynamical stability of the corresponding systems. Interestingly, the soft acoustic mode of $TiSi_2N_4$ having a frequency of ~2 THz at the K-point (shown in Fig. 2a), becomes harder (increase in the frequency at same wave-vector) when $TiSi_2N_4$ is brought in combination with any $XSi_2N_4$ monolayer. This makes $TiSi_2N_4$ easier to synthesize in combination with other compounds. For $TiSi_2N_4/MoSi_2N_4$-LH and $TiSi_2N_4/MoSi_2N_4$-VH, the hardening of this mode can be seen in Fig. 3a and 3b, respectively. The hardening of ~2 THz acoustic mode occurring at the K-point can be ascribed to the enhanced coupling at this wave-vector [62, 63].

*Electronic properties*

To understand the contribution of different orbitals from various elements that characterize our systems' electronic attributes, the total density of states (TDOS) and partial density of states (PDOS) are computed. Figure 4 illustrates the TDOS and PDOS for the respective $XSi_2N_4$ monolayers. For the monolayer $TiSi_2N_4$, the valence band maximum (VBM) is mainly dominated by N atoms '$p$' orbitals with slight contribution from Ti-$d$ orbitals, while the conduction band minima (CBM) is principally originated by the Ti-$d$ orbital, as illustrated in Fig. 4a. In the case of $MoSi_2N_4$, and $WSi_2N_4$, however, both the VBMs and CBMs maximally arise due to the Mo-/W-$d$ orbitals, which are located in the middle of each monolayer structure (Fig. 4b, c).

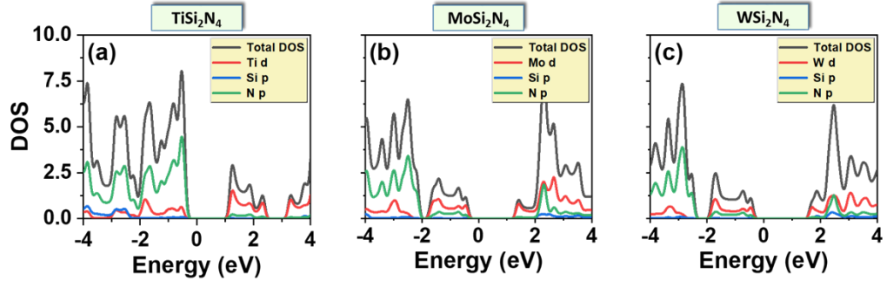

**Figure 4** Calculated TDOS and PDOS of (a) TiSi$_2$N$_4$, (b) MoSi$_2$N$_4$ and (c) WSi$_2$N$_4$ monolayers. The Fermi-level is set to zero energy.

In Fig. 5, we investigate the TDOS and PDOS for the heterostructures including TiSi$_2$N$_4$/MoSi$_2$N$_4$-LH, TiSi$_2$N$_4$/MoSi$_2$N$_4$-VH, MoSi$_2$N$_4$/WSi$_2$N$_4$-LH, and MoSi$_2$N$_4$/WSi$_2$N$_4$-VH. Figure 5(a), exhibits that the band gap decreases to a very small value of 0.137 eV for TiSi$_2$N$_4$/MoSi$_2$N$_4$-LH whereby the VBM is dominated by Mo-*d* orbitals with little contributions from N-*p* and Ti-d orbitals, whereas the CBM is maximally originated by N-*p* and Ti-*d* orbitals. Interestingly, the TiSi$_2$N$_4$/MoSi$_2$N$_4$-VH shows metallicity thereby appearing a considerable DOS at the Fermi level as shown in Fig. 5b. This metallic behavior of the VH is mainly contributed by Ti '*d*' orbitals and partly by Mo '*d*' and N '*p*' orbitals. Differently from these two heterostructures of TiSi$_2$N$_4$/MoSi$_2$N$_4$, the lateral and vertical heterostructures of MoSi$_2$N$_4$/WSi$_2$N$_4$ manifest large band gaps as shown in Fig. 5c and 5d. For MoSi$_2$N$_4$/WSi$_2$N$_4$-LH, both the VBM and CBM are dominated mainly by the '*d*' orbitals of Mo and W. Similarly, the VBM of MoSi$_2$N$_4$/WSi$_2$N$_4$-VH maximally arise due to the '*d*' orbitals of Mo and W, while the CBM is contributed by Mo '*d*' orbitals only. In conclusion, all the structures including XSi$_2$N$_4$ materials and their heterostructures reveal trivial insulating behavior, except the VH of TiSi$_2$N$_4$/MoSi$_2$N$_4$, which is metallic.

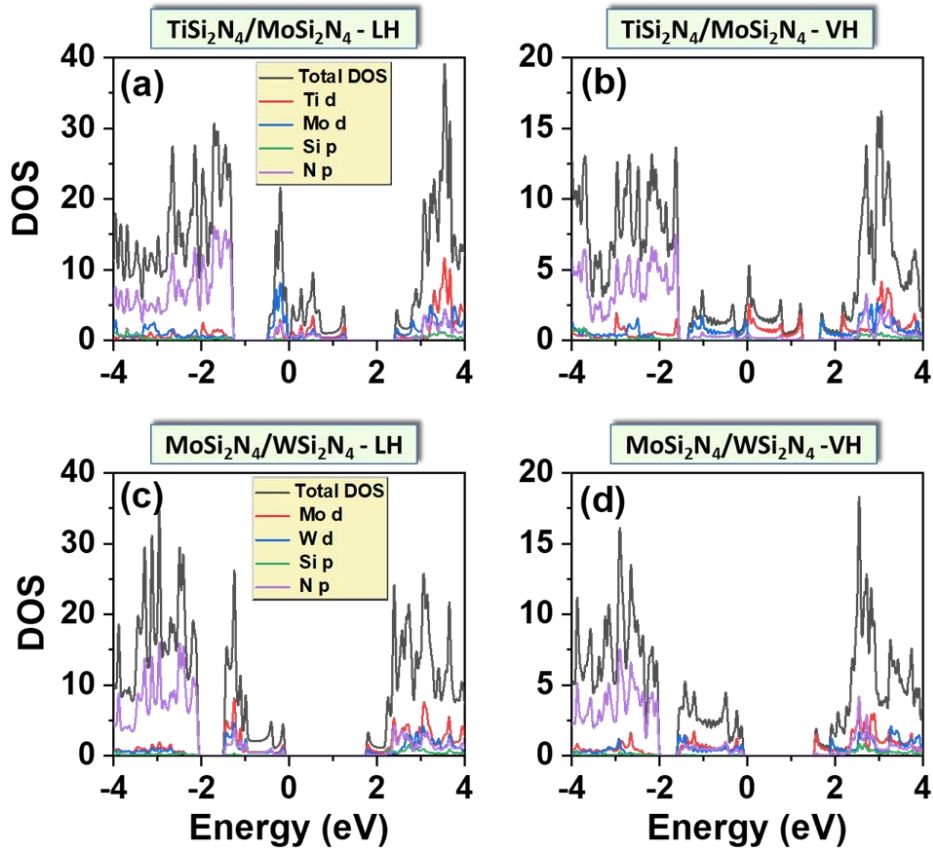

**Figure 5** Calculated TDOS and PDOS of (a) TiSi$_2$N$_4$/MoSi$_2$N$_4$-LH, (b) TiSi$_2$N$_4$/MoSi$_2$N$_4$-VH, (c) MoSi$_2$N$_4$/WSi$_2$N$_4$-LH, and (d) MoSi$_2$N$_4$/WSi$_2$N$_4$-VH. The Fermi-level is set to zero energy.

The electronic band structures were calculated to further investigate the electronic properties of XSi$_2$N$_4$ (X=Ti, Mo, W) monolayers and their heterostructures. Since the electronic band gaps calculated by DFT are usually in poor agreement with the experiments and mostly underestimated [64-66], therefore, the Hybrid functional (HSE06) [50] is applied to approximate the electronic band structures. Figure 6 demonstrates the band structures of XSi$_2$N$_4$ materials and heterostructures computed through both PBE and HSE06 potentials. The XSi$_2$N$_4$ (X=Ti, Mo, W) monolayers possess an indirect band gap of 2.61 eV, 2.35 eV, and 2.69 eV for TiSi$_2$N$_4$, MoSi$_2$N$_4$, and WSi$_2$N$_4$, respectively, as shown in Fig. 6a, b, c. For the TiSi$_2$N$_4$/MoSi$_2$N$_4$-LH, the maximum of the valence band (VBM) appears at the M-point, while the minimum of the conduction band

(CBM) occurs at the Γ-point (Fig. 6d). On the other hand, in the cases of MoSi$_2$N$_4$/WSi$_2$N$_4$-LH and MoSi$_2$N$_4$/WSi$_2$N$_4$-VH, the VBM emerges at Γ-point, whereas the CBM occurs at the K-point, as can be seen in Fig. 6e, f, suggesting that the indirect band gap nature still prevails. The values of the indirect band gaps are determined to be ~0.30, 2.34, and 2.15 eV for the TiSi$_2$N$_4$/MoSi$_2$N$_4$-LH, MoSi$_2$N$_4$/WSi$_2$N$_4$-LH, and MoSi$_2$N$_4$/WSi$_2$N$_4$-VH, respectively, which could make them promising for the application in solar cells and optoelectronic devices [67-70]. The calculated band gap values are tabulated in Table 1, which are in agreement with the available literature [27, 31, 33]. A slight difference in band gaps may arise due to the difference in lattice parameters.

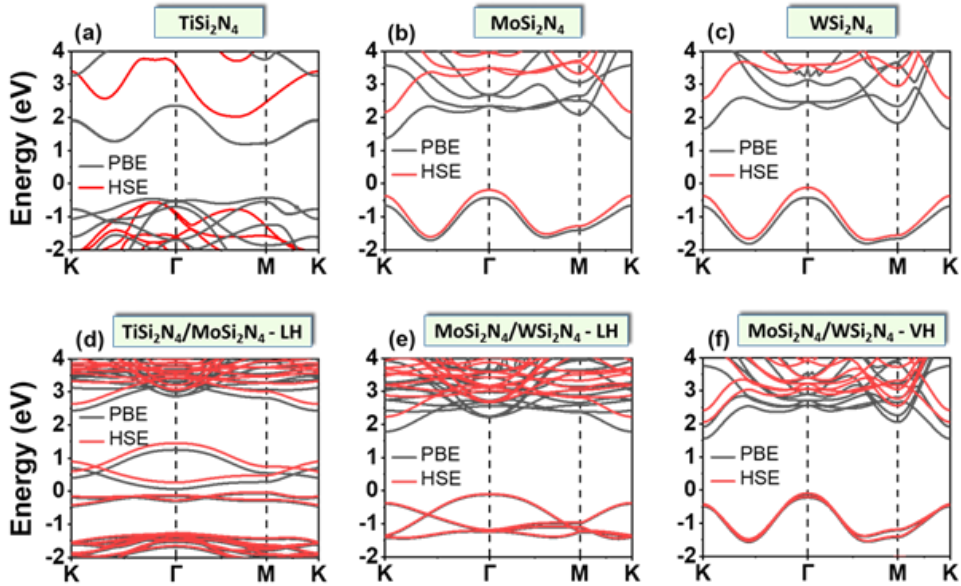

**Figure 6** Electronic band structures calculated through PBE (black color) and HSE06 (red color) for (a) TiSi$_2$N$_4$, (b) MoSi$_2$N$_4$, (c) WSi$_2$N$_4$, (d) TiSi$_2$N$_4$/MoSi$_2$N$_4$-LH, (e) MoSi$_2$N$_4$/WSi$_2$N$_4$-LH and (f) MoSi$_2$N$_4$/WSi$_2$N$_4$-VH.

**Table 1** Calculated electronic band gaps of XSi$_2$N$_4$ materials and their lateral and vertical heterostructures using PBE and HSE06.

| Material | PBE Band gap (eV) | HSE Band gap (eV) |
|---|---|---|
| TiSi$_2$N$_4$ | 1.63 | 2.40 |
| MoSi$_2$N$_4$ | 1.74 | 2.35 |
| WSi$_2$N$_4$ | 2.12 | 2.60 |
| TiSi$_2$N$_4$/ MoSi$_2$N$_4$ – LH | 0.14 | 0.30 |
| MoSi$_2$N$_4$/WSi$_2$N$_4$ – LH | 1.90 | 2.34 |
| MoSi$_2$N$_4$/WSi$_2$N$_4$ – VH | 1.70 | 2.15 |

*Thermal properties*

Based on the stable phonon dispersions of XSi$_2$N$_4$ (X=Ti, Mo, W) monolayers and their heterostructures, the thermal characteristics including entropy (S), Helmholtz free energy (F), and heat capacity $C_v$ are calculated as shown in Figure 7. The entropy of all the systems rises steadily with the temperature, consistently with the fact that increasing the temperature gives rise to several microstates in the system W, which results in an increase in the entropy logarithmically according to the Boltzmann's entropy formula (S = k$_B$logW, where $k_B$ is the Boltzmann constant). On the other hand, the Helmholtz free energy, F, is observed to decline with temperature, as shown in Fig. 7b and 7e, which agrees with $F(T) = U(T) - TS(T)$; $U(T)$ is the lattice internal energy. Furthermore, we investigate the heat capacity $C_v$ of XSi$_2$N$_4$ (X=Ti, Mo, W) monolayers and their heterostructures as a function of temperature as illustrated in Fig. 7c and 7f. Comparable heat capacities of TiSi$_2$N$_4$, MoSi$_2$N$_4$ and WSi$_2$N$_4$ are observed as reported in Fig. 7c and 7f.

Intriguingly, the heat capacity increases for the VHs and further enhances for the LHs. At room temperature, the $C_v$ values are estimated to be 106.1, 100.5, 100.8, 206.5, 415.6, 201.3 and 402.3 JK$^{-1}$mol$^{-1}$ for TiSi$_2$N$_4$, MoSi$_2$N$_4$, WSi$_2$N$_4$, TiSi$_2$N$_4$/MoSi$_2$N$_4$-VH, TiSi$_2$N$_4$/MoSi$_2$N$_4$-LH, MoSi$_2$N$_4$/WSi$_2$N$_4$-VH and MoSi$_2$N$_4$/WSi$_2$N$_4$-LH, respectively. Our estimated values of $C_v$ are relatively higher than PbSe and PbTe [71, 72] (which are thought to be excellent thermoelectric materials), suggesting the better ability of XSi$_2$N$_4$ (X=Ti, Mo, W) monolayers and their heterostructures to retain heat. The heat capacity of XSi$_2$N$_4$ (X=Ti, Mo, W) monolayers takes the lead to approach the Dulong-petit asymptotic value ( i.e. $C_v = 3NR$, where N is the number of atoms and R is the universal gas constant) at around T ~560 K. Whereas, the VHs of TiSi$_2$N$_4$/MoSi$_2$N$_4$, and MoSi$_2$N$_4$/WSi$_2$N$_4$ approach this asymptote at a relatively higher temperature, near T ~650 K, and LHs of TiSi$_2$N$_4$/MoSi$_2$N$_4$, and MoSi$_2$N$_4$/WSi$_2$N$_4$ required a higher temperature of around T~750 K and T~850K to approach this value. From this investigation, we found that $C_v$ is consistent with the entropy we computed here since the variation of entropy corresponds to the integral of ($C_v/T$) over a given temperature range.

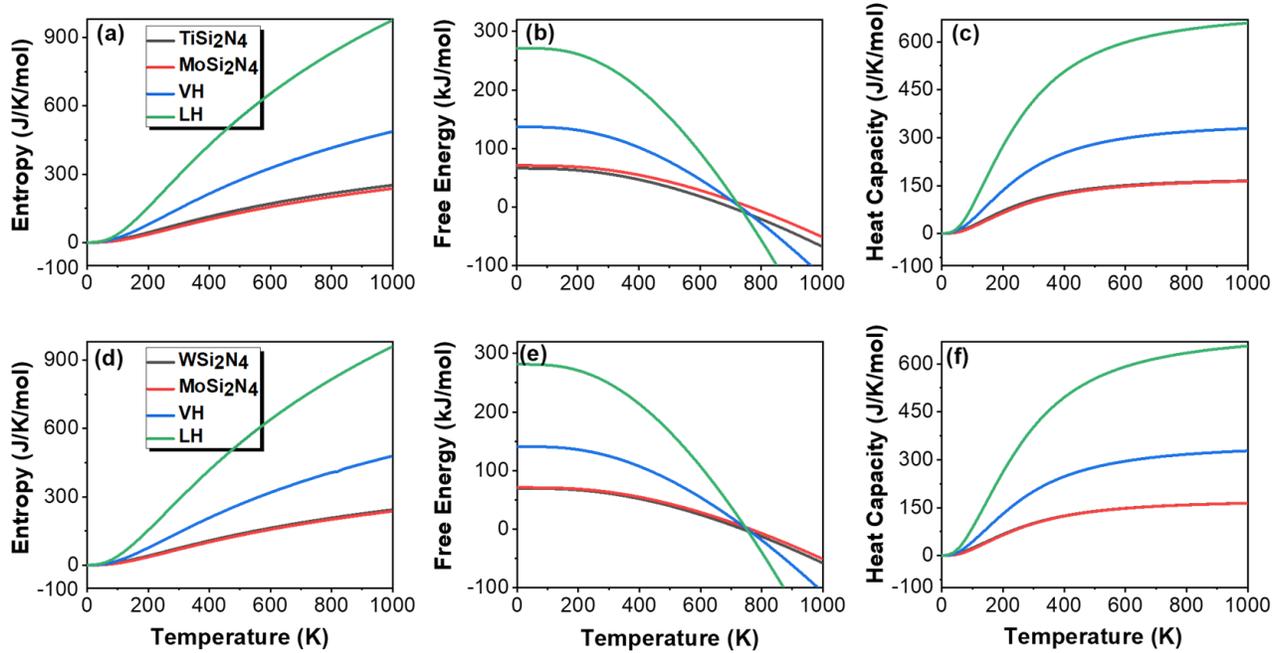

**Figure 7** Thermal properties such as entropy, free energy and heat capacity. The top panels (a), (b) and (c) represent $TiSi_2N_4$, $MoSi_2N_4$ and their lateral and vertical heterostructures, while the lower panel characterizes $MoSi_2N_4$, $WSi_2N_4$, and their lateral and vertical heterostructures.

4. Conclusions

The structural stability, electronic properties and thermal characteristics of $XSi_2N_4$ (X= Ti, Mo, W) monolayers and their lateral and vertical heterostructures are explored using density functional theory. We noticed that interfacing two different 2D structures either laterally or vertically, significantly tunes the physical properties of the parent 2D monolayers that may have technological impact in many potential applications. The structural stability of these systems was confirmed by the cohesive energies which are based on the total ground state energies and the phonon dispersions revealing the dynamical stability. Apart from their stability, our results illustrated that the lateral and vertical heterostructures present outstanding electronic and thermal features. All the three monolayers i.e. $TiSi_2N_4$, $MoSi_2N_4$, and $WSi_2N_4$ revealed indirect band gaps. Remarkably, a semiconducting to metallic transition is observed in the VH of $TiSi_2N_4/MoSi_2N_4$, while the LH of TiSi$_2$N$_4$/MoSi$_2$N$_4$ manifested a small band gap. Both the lateral and vertical heterostructures of MoSi$_2$N$_4$/WSi$_2$N$_4$ revealed semiconducting nature with modifications in their band gaps with respect to the parent monolayers. Compared to the transition metal dichalcogenides, the monolayers XSi$_2$N$_4$ and their heterostructures bear the highest values for the free energy and heat capacity. The study suggests that both the XSi$_2$N$_4$ 2D monolayers and their heterostructures could be promising in the next-generation optoelectronics and nanoelectronics.

**Acknowledgments**

This work is supported by the Foundation for Polish Science through the international research agendas program co-financed by the European Union within the smart growth operational program. We acknowledge the access to the computing facilities of the Interdisciplinary Center of Modeling at the University of Warsaw, Grants No. GB84-0 and No. GB84-7. The authors extend their appreciation to Riphah International University for funding this work under project number R-ORIC-21/FEAS-10.